\documentclass[twocolumn,showpacs,preprintnumbers,amsmath,amssymb]{revtex4}
\usepackage{graphicx}
\usepackage{dcolumn}
\usepackage{bm}
\begin{document}

\preprint{APS/123-QED}

\title{Two-magnon Raman scattering from the Cu$_{3}$O$_{4}$ layers in (Sr$_{2}$,Ba$_{2}$)Cu$_{3}$O$_{4}$Cl$_{2}$}
\author{Joakim Holmlund$^{1}$, Christopher S. Knee$^{1,4}$, Jakob Andreasson$^{1,5}$, Mats Granath$^{2}$, A. P. Litvinchuk$^{3}$, Lars B\"orjesson$^{1}$}
\affiliation{$^{1}$Department of Applied Physics, Chalmers University of Technology, SE-412 96 G\"oteborg Sweden}
\affiliation{$^{2}$Department of Physics, University of Gothenburg, SE-412 96 G\"oteborg Sweden}
\affiliation{$^{3}$Texas Center for Superconductivity and Department of Physics, University of Houston, Texas, USA}
\affiliation{$^{4}$Present address: Department of Chemistry, University of Gothenburg, SE-412 96 G\"oteborg Sweden}
\affiliation{$^{5}$Present address: Department of Cell and Molecular Biology, Uppsala University, Uppsala, Sweden}

%
%
%
%

\date{\today}
%
\begin{abstract}
(Sr$_{2}$,Ba$_{2}$)Cu$_{3}$O$_{4}$Cl$_{2}$ are antiferromagnetic insulators which are akin to the parent compounds of the cuprate superconductors but with two distinct magnetic ordering temperatures related to two magnetic Cu$_{I}$ and Cu$_{II}$ spin sublattices.
Here we present a study of these materials by means of Raman spectroscopy. Following the temperature and polarization dependence of the
data we readily identify two distinct features at around 3000 cm$^{-1}$ and 300 cm$^{-1}$  that are related to two-magnon scattering from  the two sublattices. 
The estimated spin-exchange coupling constants for the Cu$_{I}$ and Cu$_{II}$ sublattices are found to be J$_{I}\sim$139-143(132-136) meV and J$_{II}\sim$14(11) meV for Sr(Ba) compounds.
Moreover, we observe modes at around 480 and 445 cm$^{-1}$ for the Sr and Ba containing samples respectively, that disappears at the ordering temperature of the Cu$_{II}$.
We argue that this modes may also be of magnetic origin and possibly related  
to interband transitions between the Cu$_{I}$-Cu$_{II}$ sublattices.
\end{abstract}
\maketitle
\section{\label{sec:level1}Introduction}
The enormous effort of trying to understand the origin of high temperature superconductivity (HTS) in the cuprate superconductors during the past decades has led to the exploration of many new avenues regarding the field of strongly correlated electrons. It is clear that the copper oxide plane substructures, which is common to all these materials, is an essential
ingredient. A defining feature of the superconducting (SC) planes is their magnetism, and the undoped planes are well described as square lattice quantum
Heisenberg antiferromagnets (SLQHA). It is generally believed that the
magnetism has a close connection to the HTS\cite{Anderson87}\cite{scalapino:95}\cite{dahm:95}. Therefore the study of the SLQHA,
as well as its generalizations, is an important problem.
 The oxychloride 2342 phases
(Ba$_{2}$,Sr$_{2}$)Cu$_{3}$O$_{4}$Cl$_{2}$ resemble the HTS insulating parent compounds but with an additional twist of having two intercalated SLQHA (Fig. 1).  The crystal structure remains tetragonal in the interval 15 $<$ T $<$ 550 K \cite{PhysRevLett.78.535}.
The key feature of the structure is the presence of Cu$_{3}$O$_{4}$ planes, consisting of a Cu$_{I}$ sub-lattice, which is isostructural to that of the
ordinary CuO$_{2}$ planes, combined with an interpenetrating Cu$_{II}$ sub-lattice. The AFM ordering of the Cu$_{I}$ and Cu$_{II}$ spins upon lowering the temperature was
first observed from magnetic susceptibility and electron paramagnetic resonance measurements \cite{noro:90}\cite{Ramazanoglu:06}, the results indicated N\'eel
temperatures are T$_{NI}\sim$ 386 (332) K and T$_{NII}\sim$ 40 (31)K for Sr$_{2}$(Ba$_{2}$)Cu$_{3}$O$_{4}$Cl$_{2}$, respectively. The results are in line with expectations in which
the vanishing mean field from the Cu$_{I}$ spins cause an independent ordering of the Cu$_{II}$ spins but with a preferred collinearity due to quantum spin-wave
interactions \cite{shender:82}.
\\

\begin{figure}
\includegraphics[width=6cm]{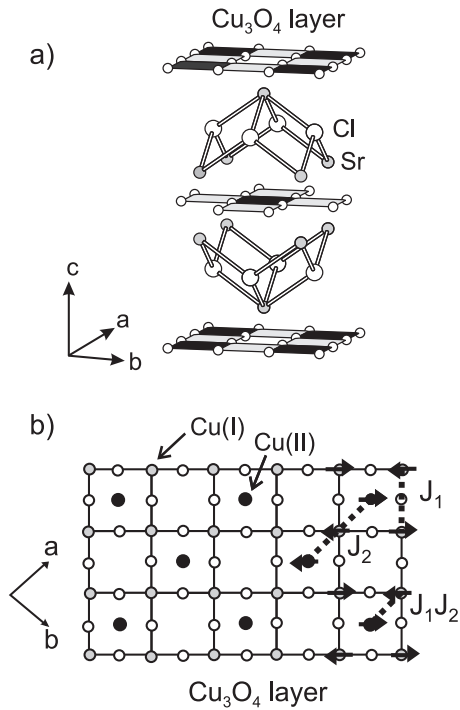}
\caption{\label{fig:struktur}(a) Crystal structure of
(Sr$_{2}$,Ba$_{2}$)Cu$_{3}$O$_{4}$Cl$_{2}$, in the Cu$_{3}$O$_{4}$ sheets shaded
square planes have a central Cu(I) ion, black square planes have
central Cu(II) ions and small open spheres are oxygens. (b)
Cu$_{3}$O$_{4}$ plane with exchange interactions between spins
indicated. Ordered spin directions for copper spins are shown as
arrows.  }
\end{figure}

From related cuprates such as YBa$_{2}$Cu$_{3}$O$_{6.1}$ and Sr$_{2}$CuO$_{2}$Cl$_{2}$ it has been shown that the magnetic exchange coupling in the Cu$_{I}$
SLQHA have values of approximately J $\sim$ 130 meV  \cite{blumberg:96}  \cite{kim:99}. For the 2342 phases there are two more couplings to consider, we have
the ordering of the second sub-lattice and its exchange coupling $J_{II}$ and also the interaction between the two magnetic lattices $J_{I-II}$(see figure
\ref{fig:struktur}b). The $J_{I-II}$ coupling is expected to be ferromagnetic by the Kanamori-Goodenough-Anderson rule due to the Cu-O-Cu 90$^{\circ}$ bond,
however, because of the vanishing mean-field for antiferromagnetically ordered Cu$_{I}$ it only comes into play through fluctuations.
Magnetization measurements \cite{chou:97} \cite{kastner:99} and comprehensive elastic, quasi elastic and inelastic neutron scattering
experiments \cite{kim:99} \cite{kim:01} show that the copper lattices display differing ordering criticalities with the Cu$_{II}$ lattice exhibiting a
2D Ising dependence. Kim et al. \cite{kim:99} \cite{kim:01} also report evidence of coupling between Cu$_{I}$ and Cu$_{II}$ as the Cu$_{I}$ out of plane gap
increases below T$_{NII}$ and state that the $J_{II}$ and $J_{I-II}$ coupling should be of similar size, around 10 meV.  In contradiction to the experimentally
determined values of $J_{I-II}$ coupling from Kim et al., Yaresko et al. \cite{yaresko:02} calculated the exchange integrals of Ba$_{2}$Cu$_{3}$O$_{4}$Cl$_{2}$
through a local density approximation including on-site correlations (LDA +U) method and reported higher values of the $J_{I-II}$ $\sim$ 20 meV. They also report slightly lower values of the $J_{II}$ coupling.
\\
When it comes to the fluctuations in this system that drive the order-from-disorder phase transitions the thermal fluctuations and quantum fluctuations do not
compete. However, introducing chemical disorder in the system will cause quenched random exchange fields that will compete with quantum fluctuations
and give rise to new interesting perspectives. Recent neutron scattering measurements by Ramazanoglu et al. \cite{Ramazanoglu:06} indicate spin glass formation
 between the two magnetic phase transitions in Co substituted  Ba$_{2}$Cu$_{2.95}$Co$_{0.05}$O$_{4}$Cl$_{2}$. Magnetization data also shows a strong
ferromagnetic enhancement upon Co substitution.\\
%
%

In this study we investigate the two interpenetrating
magnetic sublattices using Raman scattering spectroscopy (RS) for the first time. For the ordinary HTS materials the interaction of light with the spin
degrees of freedom \cite{fleury:68}
gives a two magnon (2M) peak in the Raman spectra of a frequency shift $\omega$ near 2.8 $J$ in the B$_{1g}$ scattering geometry for a tetragonal D$_{4\emph{h}}$ symmetry \cite{elliott:69} \cite{blumberg:96}. The main objective of the present study is to identify two-magnon peaks associated with the Cu$_{I}$ and Cu$_{II}$ sub-lattice ordering and determine the magnitude of the corresponding exchange interactions.
Because of the extended unit cell (containing in the plane one Cu$_{II}$
and two Cu$_{I}$ ions plus oxygens) the lattice vectors are rotated by 45$^o$ with the Cu$_{I}$ mode expected in the B$_{2g}$ and A$_{1g}$+B$_{2g}$ geometry and the Cu$_{II}$ mode
in B$_{1g}$ and A$_{1g}$+B$_{1g}$ geometry. We do indeed find two features with the expected polarization dependence at a shift of around 3000cm$^{-1}$ and 300cm$^{-1}$ respectively.
The one at low energy appears around the lower N\'eel temperature as expected for magnons of the Cu$_{II}$ antiferromagnetic order.
In addition we find an unexpected feature with a non-trivial temperature dependence around 480cm$^{-1}$ that we tentatively
assign to have a magnetic origin most likely related to
Cu$_{I}$-Cu$_{II}$ particle-hole excitations decaying through a Cu$_{I}$ two-magnon process. The Raman signature, if any, of Cu$_{I}$-Cu$_{II}$
excitations for intermediate temperatures $T_{II}<T<T_{I}$ is a challenging open problem which is complicated by the fact that the Cu$_{II}$ spins are
paramagnetic (in fact very weakly ferromagnetic\cite{chou:97}) and so cannot be treated in mean-field theory as for the Cu$_{I}$.\cite{chubukov:951} \cite{chubukov:95}

%
\section{\label{sec:level2}Experimental}
Small crystallites of Sr$_{2}$Cu$_{3}$O$_{4}$Cl$_{2}$ were obtained
by slow cooling (2.0 $^{\circ}$C/h) a partial melt of
Sr$_{2}$Cu$_{3}$O$_{4}$Cl$_{2}$ powder from 1005 $^{\circ}$C. For
Ba$_{2}$Cu$_{3}$O$_{4}$Cl$_{2}$ the melt was cooled from a lower
temperature of 965 $^{\circ}$C. The Raman measurements were
performed in back-scattering geometry using a DILOR-XY800
spectrometer/Raman microscope equipped with notch-filters and
operated in the single grating mode. The spectral resolution was
$\sim$2 cm$^{-1}$ and the diameter of the probed area was  $\sim$ 2
$\mu$m. An Ar$^{+}$/Kr$^{+}$ laser was used for excitation with a
laser power at the sample kept at 1 mW to avoid laser heating. The
wavelength used was 514.5(2.4 eV). The crystals were probed in
different scattering geometries.  To denote the scattering
geometries in our Raman experiments we use the Porto notation,
i.e., $d_{1}$(p$_{1}$p$_{2}$)d$_{2}$. Here d$_{1}$ and d$_{2}$ are
the direction of the incoming and detected light respectively and
p$_{1}$ and p$_{2}$ are the polarization of the incoming and
detected light respectively. The Porto   labels are to be taken
along the crystal axes, i.e. x=a-axis, y= b-axis and z= c-axis. For
a D$_{4\emph{h}}$ group the Raman active representations are
A$_{1g}$, B$_{1g}$, B$_{2g}$, and E$_{g}$ only while A$_{2g}$ is
silent. The specific scattering configurations presented here were
$z(x'y')\overline{z}$, $z(xy)\overline{z}$, $z(x'x')\overline{z}$,
$z(yy)\overline{z}$ and $z(xx)\overline{z}$ which correspond to
B$_{1g}$, B$_{2g}$, A$_{1g}$+B$_{2g}$,  A$_{1g}$+B$_{1g}$ and
A$_{1g}$+B$_{1g}$ Raman active representations respectively.
Variable temperature measurements were performed using a cold-finger
LHe cryostat equipped with a heater and Raman intensity calibration
was done using a BaF crystal as a standard.  All extracted
information (such as positions, intensities and shifts) from the
peaks comes from applying Gaussian fits to the two-magnon peaks and
Lorenzian fits to the phonons and then compensating for the
respective Bose occupation factor involved.

\section{\label{sec:level3}Results}
The Raman scattering from our samples will be covered in three
different subsections for Sr$_{2}$Cu$_{3}$O$_{4}$Cl$_{2}$. First we
will discuss the region around 3000 cm$^{-1}$ where we observe a strong two-magnon peak,
followed by the range around 300 cm$^{-1}$ where we observe another two-magnon excitation. Finally, we will
discuss the phonon spectra and a feature, at around 480
cm$^{-1}$. We also present data from measurements
of Ba$_{2}$Cu$_{3}$O$_{4}$Cl$_{2}$ as a comparison.
\begin{figure}[!h]
\includegraphics[width=9cm]{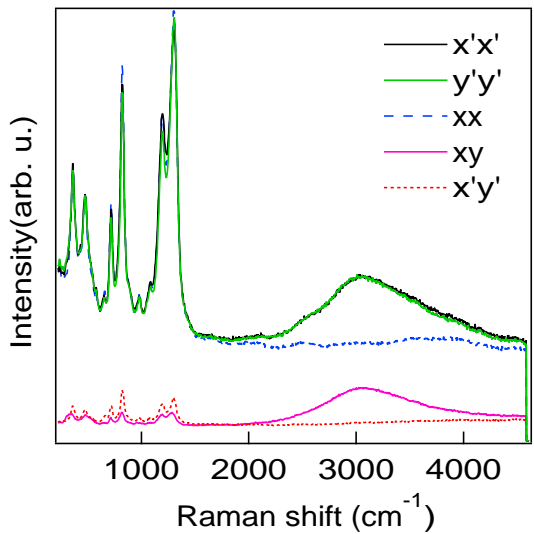}
\caption{\label{selectionrules} (Color online)Raman spectra for
Sr$_{2}$Cu$_{3}$O$_{4}$Cl$_{2}$ in $z(x'y')\overline{z}$,
$z(x'x')\overline{z}$, $z(xx)\overline{z}$, $z(xy)\overline{z}$ and
$z(y'y')\overline{z}$ scattering configuration. The spectra are
collected with an excitation wavelength of 514.5 nm at RT.}
\end{figure}
\begin{figure}[!h]
\includegraphics[width=9cm,height=8cm]{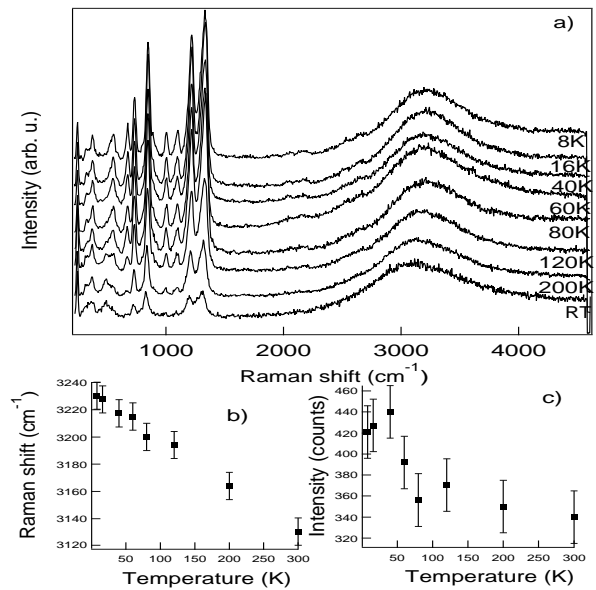}
\caption{\label{tdep3000}a) Temperature dependent Raman spectra for Sr$_{2}$Cu$_{3}$O$_{4}$Cl$_{2}$ in  $z(xy)\overline{z}$ scattering configuration for excitation wavelength of 514.5 nm from RT down to 8 K. The graphs are shown with a verical offset in order to separate the spectra. b) and c) shows the position and integrated intensity of the $\sim$ 3000 cm$^{-1}$ peak plotted against the temperature respectively. The spectra are Bose-compensated.}
\end{figure}
\newpage%
\subsection{\label{sec:level4}Two-magnon peak at around 3000 cm$^{-1}$}
This feature has a strong polarization selection rule dependence.
Figure \ref{selectionrules} shows that it is predominant in
$z(xy)\overline{z}$, $z(x'x')\overline{z}$ and $z(y'y')\overline{z}$
scattering configurations. It is very weak or absent in
$z(xx)\overline{z}$ and $z(x'y')\overline{z}$ scattering
configurations. Moreover the perfectly overlapping
$z(x'x')\overline{z}$ and $z(y'y')\overline{z}$ scattering
configurations reflects the isotropic xy plane. This polarization
dependence indicates that the feature is a two-magnon peak of the
Cu$_{I}$ sublattice. In Figure \ref{tdep3000} we present a
temperature dependent study that shows how the peak hardens
continuously from around 3130 cm$^{-1}$ to 3230 cm$^{-1}$ from RT to
8K.

Both the asymmetric shape of the feature, together with the
excitation energy, further strongly support the assignment of the
peak to a two-magnon excitation from the Cu$_{I}$ magnetic
superstructure seen before in similar compounds with square Cu-O
lattice \cite{elliott:69} \cite{blumberg:96}.%

\subsection{\label{sec:level5}Two-magnon peak at around 300 cm$^{-1}$}
\begin{figure}[!h]
\includegraphics[width=9cm]{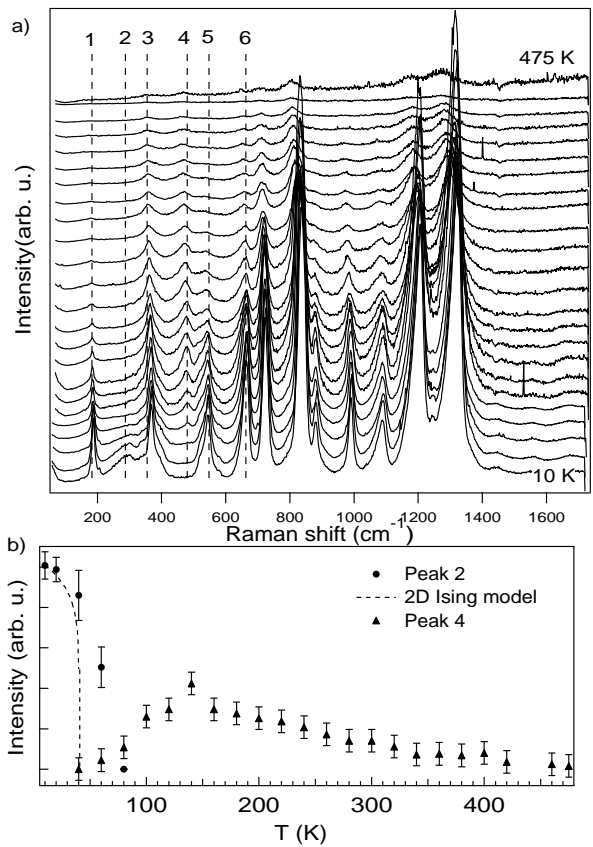}
\caption{\label{tdep300} a) Temperature dependent Raman spectra for Sr$_{2}$Cu$_{3}$O$_{4}$Cl$_{2}$ in  $z(x'y')\overline{z}$ scattering configuration for excitation wavelength of 514.5 nm from 300 down to 10 K.  The peaks numbered 1, 2, 3, 4, 5 and 6 correspond to energies of $\sim$180, 310 (2-magnon), 360, 480, 543 and 660 cm$^{-1}$ respectively (the spectra taken at 475 K is scaled by a factor of 3). b) The integrated intensities of the $\sim$310 and $\sim$480 cm$^{-1}$ modes together with a dotted curve based on a 2D Ising model (T$_{N}$-T)$^{\beta}$, where T$_{N}$=40K is the N\'eel  temperature determined from ref. \cite{noro:90} and $\beta$=0.125. All spectra are Bose--compensated.}
\end{figure}

Figure \ref{tdep300} shows the temperature dependence of the low energy region up to 1600 cm$^{-1}$, the figure shows the behavior of the phonons and magnetic interactions in the structure in $z(x'y')\overline{z}$ scattering configuration.

In order to separate the features we mark the peaks 1, 2, 3, 4, 5 and 6 which corresponds to energies of $\sim$180, 310 (two-magnon), 360, 480, 543 and 660 cm$^{-1}$ respectively.
Most interesting are peaks number two and four, which emerge and dissapear at low temperatures respectively. The behavior of the feature at around 300 cm$^{-1}$, labeled number 2, shows that the peak emerges at low temperatures, first appearing between 80 and 60K, and then the intensity grows rapidly down to 20 K before it saturates at a position $\sim$ 310 cm$^{-1}$(inset fig.\ref{tdep300}, the inset includes a dotted curve based on a 2D Ising model (T$_{N}$-T)$^{\beta}$, where T$_{N}$=40K). In contrast to the 3000 cm$^{-1}$ peak this peak is seen in $z(x'y')\overline{z}$ scattering configuration, where the former was seen in $z(xy)\overline{z}$ scattering configuration. Both its temperature and polarisation dependence indicate that this feature arises from Cu$_{II}$ 2-magnon scattering.
\subsection{\label{sec:level6}The phonon spectra including the feature at around 480 cm$^{-1}$.}
\begin{figure}[!h]
\includegraphics[width=8cm]{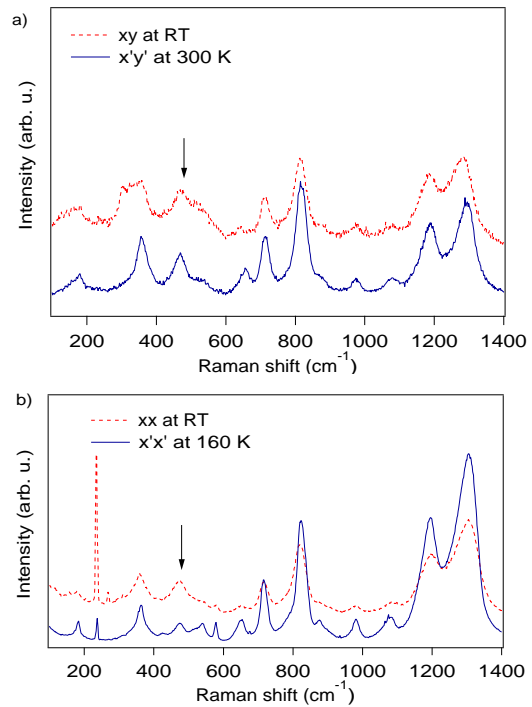}
\caption{\label{selection480} (Color online) Raman spectra for  Sr$_{2}$Cu$_{3}$O$_{4}$Cl$_{2}$ in a) $z(x'y')\overline{z}$ and $z(xy)\overline{z}$, b) $z(x'x')\overline{z}$ and $z(xx)\overline{z}$ scattering configuration. The spectra are collected with a excitation wavelength of 514.5 nm at RT, 300 K and 160K. The 480 cm$^{-1}$ peak is marked in the figures.}
\end{figure}

\begin{figure}[!h]
\includegraphics[width=8cm]{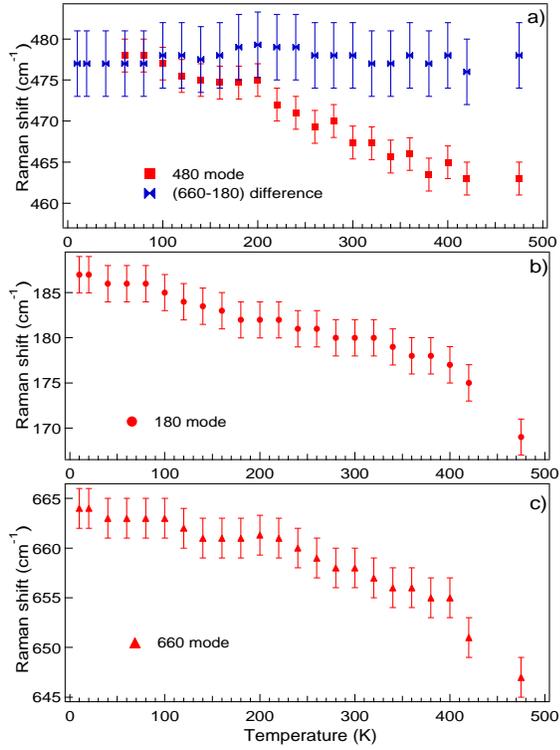}
\caption{\label{differencemode} (Color online) Temperature dependence of the Raman shifts  for the a) 480 b) 180 and c) 660  cm$^{-1}$ phonon modes in $z(x'y')\overline{z}$ scattering configuration for excitation wavelength of 514.5 nm from 475 down to 10 K, the information is extracted from applying Lorenzian fits of the three phonon modes. 
In addition a) shows the difference of the 660 and 180 cm$^{-1}$ modes with increasing temperature. }
\end{figure}

 The phonon Raman spectra of Sr$_{2}$Cu$_{3}$O$_{4}$Cl$_{2}$ is complicated in many ways. It has a number of phonons that do not follow normal Bose-Einstein
occupation statistics, instead there are several phonons, including higher order phonons that grow anomalously with decreased temperature, see figure \ref{tdep3000} and  \ref{tdep300}. This has earlier been reported for compounds including spin chains, and has been explained by Fr\"ohlich interaction induced activation of longitudinal phonons  \cite{PhysRevB.55.R8638} \cite{PhysRevB.53.R14733}\cite{holmlund:134502}. These effects have only been seen when the incoming and scattered light has been polarized along the chains and ladders. However, in (Sr$_{2}$, Ba$_{2}$)Cu$_{3}$O$_{4}$Cl$_{2}$ we observe this extreme behavior in cross-polarized scattering configuration, see figure \ref{tdep3000} and  \ref{tdep300}. In $z(x'y')
\overline{z}$ (fig. \ref{tdep300}) and $z(xy)\overline{z}$ (fig. \ref{tdep3000}) scattering configuration  we observe a peak at around 480 cm$^{-1}$ with different and even more peculiar temperature dependence.  The peak can be seen in all measured scattering configurations ($z(xy)
\overline{z}$,$z(x'y')\overline{z}$, $z(xx)\overline{z}$ and $z(x'x')\overline{z}$) See figure \ref{selection480}. Furthermore, the peak looses intensity upon lowering the temperature from 475 K and it approaches zero at a finite temperature at around 60K.
Since the crystal structure is stable in this interval, a possible explanation would be that its origin is a two-phonon difference mode.  At first sight there are two modes at around 660 cm$^{-1}$ (peak 6) and 180 cm$^{-1}$ (peak 1) that
appear to match the difference criterion as shown in figure \ref{differencemode}. Here we plot the temperature dependence of the shifts for all three modes (180, 480 and 660 cm$^{-1}$) together with the difference between the 180 and 660 cm$^{-1}$ phonon modes. The difference condition starts to break down above 200K and the gap increases at higher temperatures, displaying a clear mismatch at 475K. Hence the difference mode explanation is not likely.
Instead, given the modes sensitivity to the ordering of the Cu$_{II}$ spins, it seems more likely that the mode has a magnetic origin and if so probably related to the Cu$_{I}$ magnon which is the only expected coherent magnetic mode at these temperatures. The energyshift from the Cu$_I$ two-magnon decay may be due to the Cu$_{I}$-Cu$_{II}$ energy difference. 

\subsection{\label{sec:level7}Ba$_{2}$Cu$_{3}$O$_{4}$Cl$_{2}$}
\begin{figure}[!h]
\includegraphics[width=9cm]{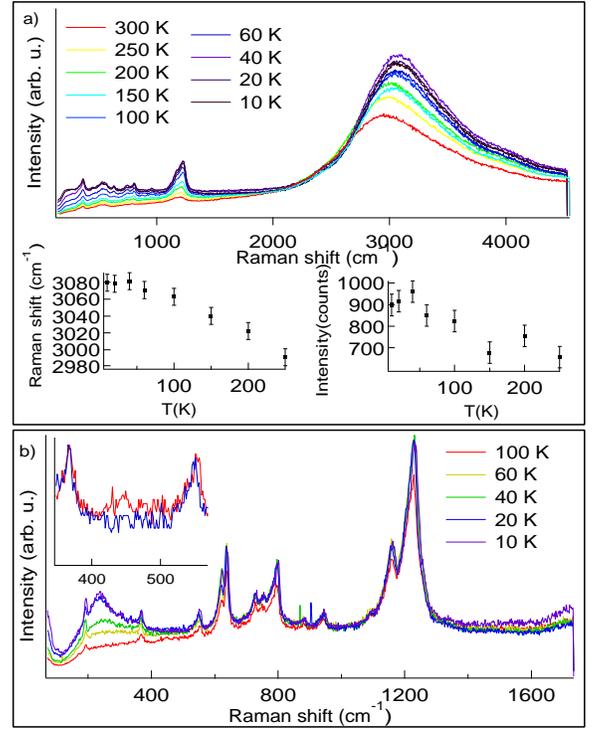}
\caption{\label{batdep} (Color online) a) Temperature dependent Raman spectra for Ba$_{2}$Cu$_{3}$O$_{4}$Cl$_{2}$ in $z(xy)\overline{z}$ scattering configuration. The left and the right inset shows the position and integrated intensity of the $\sim$ 3000 cm$^{-1}$ peak plotted against the temperature respectively.  b) Raman spectra for Ba$_{2}$Cu$_{3}$O$_{4}$Cl$_{2}$ in $z(x'y')\overline{z}$ scattering configuration for excitation wavelength of 514.5 nm from 300 to 10 K. The small inset shows a zoom up of the $\sim$445 cm$^{-1}$ mode. The spectra are Bose-compensated.}
\end{figure}

\begin{figure}[!h]
\includegraphics[width=9cm]{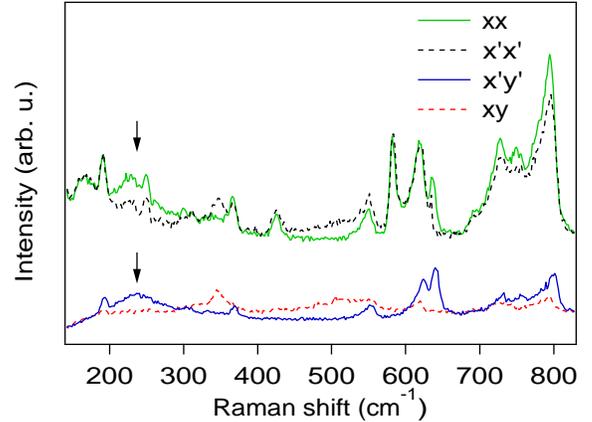}
\caption{\label{selectionrules230} (Color online) Raman spectra for  Ba$_{2}$Cu$_{3}$O$_{4}$Cl$_{2}$ in  $z(xy)\overline{z}$, $z(x'y')\overline{z}$, $z(xx)\overline{z}$ and $z(x'x')\overline{z}$ scattering configuration. The spectra are collected with an excitation wavelength of 514.5 nm at 10 K. The $\sim$230 cm$^{-1}$ peak is marked in the figures.}
\end{figure}

For comparison we made similar Raman measurements on a compound where Ba was substituted for Sr, see figure \ref{batdep}. All spectra showed a similar behavior in the Raman scattering
data as for the Sr$_{2}$Cu$_{3}$O$_{4}$Cl$_{2}$ compound, with the exception of somewhat different energies for the excitations upon switching to Ba. Figure
\ref{batdep} a) displays the temperature dependence for the $\sim$ 3000 cm$^{-1}$ peak. The peak hardens from around 2990 cm$^{-1}$ and grows in intensity
down to the second N\'eel temperature at around 30 K and then it remains at approx. 3080 cm$^{-1}$ and shows a weak drop in intensity. Figure \ref{batdep} b) displays the temperature dependence for the $\sim$ 300 cm$^{-1}$
peak. The peak is more prominent than in the Sr-substituted compound and the position at 10 K is at $\sim$ 237 cm$^{-1}$ compared with 310 cm$^{-1}$ for the
former. Figure \ref{selectionrules230} displays the low energy region in four different scattering configurations with light polarized along xx, xy, x'x' and
x'y'. From the figure one can clearly observe the peak at $\sim$230 cm$^{-1}$ in x'y' and xx. This is easily compared with the Cu$_{II}$
magnetic sublattice seen in figure \ref{fig:struktur} b). The selection rules apply well if one considers that the Cu$_{II}$ magnetic superstructure is
rotated by 45 $^{\circ}$s with respect to the Cu$_{I}$ magnetic superstructure. The shift to lower energy for the two-magnon peaks in the Ba substituted compound compared to the Sr substituted compound would also be in accordance with the expected weaker superexchange, due to increased Cu$_{I}$-Cu$_{I}$ and Cu$_{II}$-Cu$_{II}$ separations. In addition we observed the corresponding Cu$_{I-II}$magnetic mode for this compound, the position at 100 K is at $\sim$445 cm$^{-1}$ (see inset of  figure \ref{batdep} b).
\section{\label{sec:level8}Discussion}
The results clearly distinguish several interesting features in the Raman spectra, of which at least two most probably stem from magnetic scattering. This assertion is strengthened as the 2342 phases exhibit no structural transitions upon decreasing temperature \cite{PhysRevLett.78.535}. The most prominent feature is the broad and intense peak at $\sim$ 3000 cm$^{-1}$. The energy of the feature is similar to other SLQHA system with a J $\sim$ 130 meV \cite{blumberg:96} and the selection rules apply well for a typical two-magnon excitation from the Cu$_{I}$ magnetic sublattice. Table I reveals a slight shift to lower energies for the peak maxima in Ba$_{2}$Cu$_{3}$O$_{4}$Cl$_{2}$ reflecting the increased Cu$_{I}$-Cu$_{I}$ separation and weakened superexchange in this compound, i.e. the a-parameters for the Sr and Ba analogues are 5.462 and 5.517 \AA~respectively.\\\\
\begin{tabular}{|c|c|c|}
\multicolumn{3}{l}{Table I.}\\
\multicolumn{3}{l}{Peak positions and superexchange coupling constants.}\\
\multicolumn{3}{l}{Magnon peak= 2.8 J.}\\
\hline
Compound & Sr$_{2}$Cu$_{3}$O$_{4}$Cl$_{2}$ & Ba$_{2}$Cu$_{3}$O$_{4}$Cl$_{2}$\\
\hline
Cu$_{I}$ (cm$^{-1}$)&3130-3230&2990-3080\\
$J_{I}$(meV)&139-143&132-136\\
T (K)&300-8&300-10\\
\hline
Cu$_{II}$ (cm$^{-1}$)&$\sim$310&$\sim$ 240\\
$J_{II}$ (meV)&13.7&10.6\\
T (K)&10&10\\
\hline
\end{tabular}\\
\\\\
The next feature to consider is the peak that emerges at around the second N\'eel temperature $\sim$ 31 and 40 K for Ba and Sr compounds respectively \cite{noro:90}\cite{Ramazanoglu:06}. The selection rules for this feature clearly shows a relation to the Cu$_{II}$ magnetic superstructure considering a 45 $^{\circ}$ rotation of the magnetic superstructure with respect to the Cu$_{I}$ lattice. The energy of the peak also allows the J$_{II}$ coupling to be estimated and, as shown in table I, the derived values agree pretty well with measured J-coupling constants from neutron scattering experiments J $\sim$ 10 meV \cite{kim:99} \cite{kim:01}. The red-shift observed for the Ba compound also fits the expected trend. One further point concerns the apparently non 2D-Ising growth of the feature shown in the inset of figure \ref{tdep300}. This discrepancy most likely reflects the different length scales probed by the Raman and diffraction techniques. In fact, our data show that the two-magnon first appears in the 60 K data, significantly above the T$_{NII}$ of Kim et al.\cite{kim:99}, emphasising that a long range inter-planar correlation is not required for the short wavelength spin waves probed by our light scattering experiment \cite{PhysRevB.43.6224}.
\\
The peak at around 480 cm$^{-1}$ displays an unusual non monotonic temperature dependence. The simplest explanation for this feature would be that it is a two phonon difference mode. However the behavior upon increasing the temperature appears to rule out this explanation. The intensity of the peak decreases towards lower temperatures and it completely disappears below T $\approx$60 K (Figure \ref{tdep300}).
This temperature coincides very well with the growth of the Cu$_{II}$-Cu$_{II}$ two-magnon, that reflects the ordering of the second magnetic lattice. The
peak could thus be severely affected by the Cu$_{II}$-Cu$_{II}$ ordering which points towards a magnetic origin. The present Raman experiments are done in the resonant regime
in which the current operator creates particle-hole excitations across the Mott-Hubbard gap. Excitations both within the sublattices as well as between sublattices
are expected and the energy difference $3000-480\approx 2500$cm$^{-1}$ (300meV) would
then reflect the Cu$_{I}$-Cu$_{II}$ site energy difference, which is in fact roughly in-line with an LDA estimate of 380meV\cite{Rosner}.
As mentioned in the introduction this is a
particularly difficult problem in the intermediate temperature regime where the Cu$_{II}$ spins are not ordered. Whether any additional coherent magnetic response
apart from the ordinary Cu$_{I}$ two-magnon should be expected is an open problem.
\\
Finally, we discuss briefly the profile shape of the two-magnon peaks. The $\sim$ 3000 cm$^{-1}$ magnon has a typical high energy tail seen in other similar SLQHA
systems, see for example Blumberg et al. \cite{blumberg:96}. In contrast the 300 cm$^{-1}$ two-magnon appears to have a low energy tail. Theoretical work has been relatively successful in predicting the shape and polarisation dependence of the usual (Cu$_{I}$) 2-magnon (see for example reference \cite{PhysRevB.54.3468} and references therein) for layered cuprate antiferromagnets. 
However at this stage it is difficult to comment further concerning the profile since there are at least two phonons that screeen the hight energy side making it difficult to reliably deconvolute the two-magnon scattering. Furthermore, for the Ba substituted compound the notch filter cuts the low energy side and thus prevents us from analyzing the two-magnon profile at lower energies in a satisfactory way.
It is hoped that these data presented here on the 2342 systems will provide a basis for extension of the theories for quantum antiferromagnets.

\section{\label{sec:level9}summary}
In summary we have studied the magnetic scattering in (Sr$_{2}$,Ba$_{2}$)Cu$_{3}$O$_{4}$Cl$_{2}$ through the sensitive probe of magnetic Raman scattering. Based on the
temperature dependence and selection rules we assign  the peaks at $\sim$ 3000 cm$^{-1}$ and $\sim$ 300 cm$^{-1}$ to two-magnon excitations  peaks stemming from
the Cu$_{I}$ and Cu$_{II}$ sublattice respectively. The peak positions we have obtained estimates of the strength of the
J-couplings for the Cu$_{I}$-Cu$_{I}$ and Cu$_{II}$-Cu$_{II}$ interactions of around 140 and 14 meV respectively. In addition an interesting mode is observed at $\sim$ 480 cm$^{-1}$, which most probably has a magnetic origin related to the coupling between the Cu$_{I}$ and Cu$_{II}$ sublattices. 
Further work regarding an extension of existing theories for resonant magnetic Raman response is needed to describe the intermediate (T$_{NII}< $T$< T_{NI}$) temperature range of these materials. In addition, investigations with applied magnetic fields could be of high interest in order to resolve unanswered questions regarding the magnetic interactions in these materials.

\section{\label{sec:level10}Acknowledgements}
We acknowledge financial support from the Oxide research program of the Swedish foundation for strategic Research. C.S.K acknowledges support from the European Commission sixth framework programme through the Marie Curie actions.

\end{document}